\documentclass{article}
\pdfoutput=1

\usepackage{arxiv}

\usepackage[utf8]{inputenc} 
\usepackage[T1]{fontenc}    
\usepackage{hyperref}       
\usepackage{url}            
\usepackage{booktabs}       
\usepackage{amsfonts}       
\usepackage{nicefrac}       
\usepackage{microtype}      
\usepackage{lipsum}		    
\usepackage{amsmath}
\usepackage{times}
\usepackage{hyperref}
\usepackage{float}
\usepackage{enumitem}

\RequirePackage[numbers]{natbib}
\usepackage{graphicx}

\usepackage{lipsum}
\usepackage[american]{babel}
\usepackage{adjustbox}
\usepackage[figuresright]{rotating}

\usepackage{caption}
\usepackage{lineno}
\usepackage{comment}
\usepackage{xcolor}
\newenvironment{revision}{\color{black}}{\color{black}}
\DeclareCaptionType{InfoBox}

\title{Footprint of Publication Selection Bias on Meta-Analyses in Medicine, Environmental Sciences, Psychology, and Economics} 

\date{} 
\renewcommand{\headeright}{}
\renewcommand{\undertitle}{}

\author{
František Bartoš$^{1,2}$*, Maximilian Maier$^{3}$*, Eric-Jan Wagenmakers$^{1}$, \\
\textbf{Franziska Nippold$^{1}$, Hristos Doucouliagos$^{4}$, John P. A. Ioannidis$^{5,6,7,8,9}$,}\\
\textbf{Willem M. Otte$^{10}$, Martina Sladekova$^{11}$, Teshome K. Deresssa$^{12}$, Stephan B. Bruns$^{12,13}$}\\ 
\textbf{Daniele Fanelli$^{14,15}$, T.D. Stanley$^{4}$}
\\
\\
    \footnotesize{$^*$Both authors contributed equally} \\
    \footnotesize{$^1$ Department of Psychological Methods, University of Amsterdam} \\ 
    \footnotesize{$^2$ Institute of Computer Science, Czech Academy of Sciences} \\
    \footnotesize{$^3$ Department of Experimental Psychology, University College London} \\
    \footnotesize{$^4$ Department of Economics, Deakin University} \\
    \footnotesize{$^5$ Meta-Research Innovation Center at Stanford (METRICS)} \\
    \footnotesize{$^6$ Department of Epidemiology and Population Health, Stanford University School of Medicine}\\
    \footnotesize{$^7$ Stanford Prevention Research Center, Department of Medicine, Stanford University School of Medicine }\\
    \footnotesize{$^{8}$ Department of Biomedical Data Science, Stanford University School of Medicine} \\
    \footnotesize{$^{9}$ Department of Statistics, Stanford University School of Humanities and Sciences} \\
    \footnotesize{$^{10}$ Department of Pediatric Neurology, UMC Utrecht Brain Center,} \\
        \footnotesize{University Medical Center Utrecht, and Utrecht University} \\
    \footnotesize{$^{11}$ School of Psychology, University of Sussex} \\
    \footnotesize{$^{12}$ Centre for Environmental Sciences, Hasselt University} \\
    \footnotesize{$^{13}$ Department of Economics, University of G\"{o}ttingen} \\
    \footnotesize{$^{14}$ Department of Methodology, London School of Economics and Political Science} \\
    \footnotesize{$^{15}$ School of Social Sciences, Heriot-Watt University} \\
\\
\footnotesize{Correspondence concerning this article should be addressed to František Bartoš at f.bartos96@gmail.com.}\\
}

\renewcommand{\shorttitle}{Footprint of Publication Selection Bias}

\begin{document} 

\maketitle 

\begin{abstract} 
Publication selection bias undermines the systematic accumulation of evidence. To assess the extent of this problem, we survey over 68,000 meta-analyses containing over 700,000 effect size estimates from medicine \begin{revision}(67,386/597,699)\end{revision}, environmental sciences \begin{revision}(199/12,707)\end{revision}, psychology \begin{revision}(605/23,563)\end{revision}, and economics \begin{revision}(327/91,421)\end{revision}. Our results indicate that meta-analyses in economics are the most severely contaminated by publication selection bias, closely followed by meta-analyses in environmental sciences and psychology, whereas meta-analyses in medicine are contaminated the least. \begin{revision}After adjusting for publication selection bias, the median probability of the presence of an effect decreased from 99.9\% to 29.7\% in economics, from 98.9\% to 55.7\% in psychology, from 99.8\% to 70.7\% in environmental sciences, and from 38.0\% to 29.7\% in medicine.\end{revision} \begin{revision}The median absolute effect sizes (in terms of standardized mean differences) decreased from $d = 0.20$ to $d = 0.07$ in economics, from $d = 0.37$ to $d = 0.26$ in psychology, from $d = 0.62$ to $d = 0.43$ in environmental sciences, and from $d = 0.24$ to $d = 0.13$ in medicine.\end{revision}
\end{abstract}

\newpage


\section{Introduction}

Publication selection biases (PSB) are defined as the selective reporting of results in ways that deviate from the objective, complete scientific record. PSB may entail the suppression of ``negative'' findings or the conversion of ``negative'' results into more ``positive'' ones (e.g., those with more favorable $p$-values and/or with larger effect sizes) and might represent a problem in all scientific disciplines, e.g., \cite{chavalarias2010science,dwan2008systematic,rosenthal1964further,wicherts2017weak,otte2022analysis}. Studies that examine the self-reported behavior of researchers show that 78\% of researchers failed to report all dependent measures of a study \cite{john2012measuring} (however, see \cite{fiedler2016questionable}, for a response that suggests a lower proportion). Some studies also suggest that PSB might be modestly increasing in some areas, although the exact nature, prevalence, and impact of PSB is unknown and likely to be variable across scientific fields \cite{de2015surge,fanelli2017meta}.

To gauge the extent of the PSB, one would need to have access to the complete scientific record or a representative and wide-coverage sample of it. However, this is infeasible as much of the relevant data is not publicly recorded. Instead, the footprint of PSB is indirectly probed by re-analyzing meta-analyses in several specific fields with different statistical techniques
\cite{fanelli2017meta,ioannidis2017power,mathur2019finding,stanley2018what,van2019publication,schwab2021assessing} and focusing on patterns in the published results that would herald the presence of PSB. All these available methods try to identify the footprint of PSB, and thus their results need to be interpreted with caution since these patterns (e.g., correlations of effect sizes and standard errors) may sometimes be due to factors other than PSB (e.g., genuine heterogeneity across studies). However, when large numbers of meta-analyses show the same patterns, this constitutes a probable footprint of PSB, which can be used to estimate its relative magnitude across different fields.

Previous field-wide assessments of PSB suggested that the prevalence of over-reporting positive results and other possible symptoms of bias increased moving from the physical to the biological and the social sciences, and even suggested that problems might be worsening over time in the latter \cite{fanelli2017meta,kuhberger2014publication, fanelli2010positive,ioannidis2011excess, scheel2021excess, sterling1959publication, fanelli2012negative}. However, these estimates were based on proxy measures of PSB that have several limitations.

To our knowledge, no previous survey of the potential footprint of PSB has used state-of-the-art methods. Our proposed approach is more comprehensive than past surveys: employing different strategies to identify potential PSB, using new measures of PSB, and analyzing a much larger number of research studies covering the fields of medicine, environmental sciences, psychology, and economics.

\section{Methods}

\subsection{Data Sets}
\label{app:datasets}

We used five large data sets from medicine, environmental sciences, psychology, and economics. The data set from medicine comprises meta-analyses of continuous and dichotomous outcomes obtained from the Cochrane Database of Systematic Reviews published between 1997 and 2020. The data set from environmental sciences comprises the meta-analyses of mean differences, odds ratios, and correlation coefficients by Deressa and colleagues \cite{deressa2022selective} published between 2010 and 2020. The data sets from psychology comprise the meta-analyses of mean differences and correlation coefficients by Stanley and colleagues \cite{stanley2018what} published between 2011 and 2016 combined with a random sample of meta-analyses published in psychological journals by Sladekova and colleagues \cite{sladekova2022estimating} published in 2008 and 2018. Finally, the data set from economics comprises the extended data set of meta-analyses of regression and correlation coefficients by Ioannidis and colleagues \cite{ioannidis2017power} published between 1967 and 2021. Eighty-four meta-analyses were part of both the \cite{ioannidis2017power} and \cite{stanley2018what} data set. Since each of the meta-analyses could be classified in both fields (psychology or economics), we did not remove them from either of the data sets. From each data set we only used meta-analyses with at least three estimates reported using standardized effect size metrics such as log odds ratios, standardized mean differences, and (partial) correlation coefficients that can be transformed to a common standardized mean difference effect size metric, Cohen's $d$.

\subsubsection{Medicine}

The data set from medicine comprises meta-analyses of continuous and dichotomous outcomes obtained from the Cochrane Database of Systematic Reviews (CSDR) published between 1997 and 2020. We identified systematic reviews in the CDSR through PubMed, limiting the period to Jan 2000 -- May 2020. For that, we used the NCBI's EUtils API with the following query: ``Cochrane Database Syst Rev''[journal] AND (``2000/01/01''[PDAT]: ``2020/05/31''[PDAT]). For each review, we downloaded the XML meta-analysis table file (rm5-format) associated with the review's latest version. We extracted the tables with continuous and dichotomous outcomes from these rm5-files with a custom Javascript and R programs (\url{https://github.com/wmotte/cochrane2022}).

We selected meta-analysis tables based on the highest aggregation reported in the CSDR. For each meta-analysis, we removed estimates based on one or fewer participants in the control or treatment group and used all meta-analyses with at least three effect size estimates.

\subsubsection{Environmental Sciences}
The environmental sciences data set consists of meta-analyses of mean differences, correlation coefficients, and odds ratios published between 2010 and 2020. The literature search was performed in the Scopus database using the query: "TITLE-ABS-KEY (“meta analy*” OR “meta-analy*” OR “metaanaly*” OR “meta reg*” OR “meta-reg*” OR “metareg*”) AND SUBJAREA (envi)" on July 21, 2020. Detailed information about the sampling strategy and inclusion/exclusion criteria used can be found in \cite{deressa2022selective}.

\subsubsection{Psychology}

The data set from psychology comprise the data set of meta-analyses of mean differences and correlation coefficients of Stanley and colleagues \cite{stanley2018what} published between 2011 and 2016 combined with data from Sladekova and colleagues \cite{sladekova2022estimating}, a random sample of 433 meta-analyses from 90 articles published in 2008 and 2018. See \cite{stanley2018what} and \cite{sladekova2022estimating} for more details about the collected data sets. None of the meta-analyses by \cite{sladekova2022estimating} were published in Psychological Bulletin, precluding overlap with Stanley and colleagues \cite{stanley2018what} data set.

\subsubsection{Economics}

The data set from economics comprise the extended data set of meta-analyses of regression and correlation coefficients of Ioannidis and colleagues \cite{ioannidis2017power} published between 1967 and 2021. The meta-analyses were identified using various search engines (e.g.. Econlit and Scopus), publisher sites (e.g., Science Direct, Sage, and Wiley), webpages of researchers known to publish meta-analyses, and by searching all volumes of individual journals that are known to publish meta-analyses. We also emailed 109 research teams (associated with either sole authored or co-authored meta-analyses) for data, with a 67\% response rate. The search for data ended May 30th, 2021. 

We selected meta-analyses of standardized mean differences, (partial) correlation coefficients, and mean differences (if enough information was available to compute the standardized mean differences).

\subsubsection{Effect Size Calculation}

In cases where the data set did not already feature standardized effect size (Cohen's $d$, correlation coefficient $r$, log(\emph{OR}), or Fisher's $z$), we used the \texttt{metafor} \texttt{R} package \cite{metafor} to calculate the standardized effect sizes. For dichotomous outcomes with zero cell counts, we used the default empty cell correction, adding $1/2$ to empty cells. Finally, we converted all standardized effect sizes to Fisher's $z$ by using the formulas in \cite{borenstein2009introduction}.

\subsection{Publication Bias Adjustment \begin{revision}
with Bayesian Model-Averaging \end{revision}}
\label{app:methods}
We used the PSB detection and correction technique RoBMA-PSMA \cite{maier2020robust, bartos2021no}. \begin{revision}RoBMA employs Bayesian model-averaging \cite{hoeting1999bayesian, fragoso2018bayesian} and\end{revision} combines the best of two well-performing publication bias adjustment methods: selection models \begin{revision} with six different weight functions that adjust for publication selection across a combination of statistical significance and direction of the effect\end{revision} \cite{vevea1995general} and PET-PEESE \begin{revision}, which adjusts for the relationship between effect sizes and standard errors or standard errors squared\end{revision} \cite{stanley2017finding}. \begin{revision} Bayesian model-averaging allows us to combine these publication bias adjustment methods based on their predictive adequacy, such that models that predict well have a larger impact on the inference. In that way, we can evaluate the evidence in favor or against the hypothesis of PSB and its impact without committing to any single estimation or correction method \cite{ hoeting1999bayesian}. \end{revision} 

\begin{revision}We used the default RoBMA parameterization which was shown to achieve better performance in both simulation studies and real data examples than either of publication bias adjustment methods alone \cite{bartos2021no}. It gives equal prior model probabilities to models assuming the presence vs. absence of an effect, heterogeneity, and publication selection bias. RoBMA employs a standard normal distribution on the effect size, $\mu \sim \text{Normal}(0, 1)$, empirically informed Inverse-gamma distribution on the heterogeneity, $\tau \sim \text{Inverse-Gamma}(1, 0.15)$ \cite{vanerp2017estimates}, cumulative unit Dirichlet prior distributions on publication probabilities, and Cauchy prior distributions on the PET-PEESE regression coefficients, $\text{PET} \sim \text{Cauchy}_+(0, 1)$, $\text{PEESE} \sim \text{Cauchy}_+(0, 5)$.\end{revision}

\begin{InfoBox}[H]
\caption{Bayes factors}
\fbox{
\begin{minipage}{\textwidth}
\small

The Bayes factor is the key inference criterion for much of Bayesian statistics, e.g., \cite{wrinch1921on, kass1995bayes}. It compares the relative predictive accuracy (i.e., likelihood of the data) under competing hypotheses (e.g., $\mathcal{H}_1$ vs.  $\mathcal{H}_0$) and it can also be expressed as the ratio of prior and posterior model odds, 
\begin{equation*} \label{eq:BF}
\underbrace{ \frac{p(\text{data} \mid \mathcal{H}_1)}{p(\text{data} \mid \mathcal{H}_0)}}_{\substack{\text{Bayes factor}}} = \underbrace{ \frac{p(\mathcal{H}_1 \mid \text{data})}{p(\mathcal{H}_0 \mid \text{data})}}_{\substack{\text{Posterior odds}}} \,\, \bigg/ \underbrace{ \frac{p(\mathcal{H}_1)}{p(\mathcal{H}_0)}}_{\substack{\text{Prior odds}}}.
\end{equation*}

Although the Bayes factor is a continuous measure of strength of evidence, the following rules of thumb may aid interpretation: Bayes factors between 1 and 3 are commonly regarded as weak evidence, Bayes factors between 3 and 10 as moderate evidence, and Bayes factors larger 10 as strong evidence for the alternative (or the hypothesis at the top of Equation \ref{eq:BF}). When the evidence for the null is considered, the Bayes factor is simply inverted. In other words, a Bayes factor between 1/3 and 1 is considered weak evidence, a Bayes factor between 1/10 and 1/3 moderate and smaller 1/10 strong evidence for the null (e.g., \cite{jeffreys1939theory}; \cite{lee2013bayesian}).
\end{minipage}
}
\label{box:Bayesfactor}
\end{InfoBox}

\subsection{Measures}
\label{app:measures}
 \begin{revision}For each meta-analysis, we used RoBMA to calculate the (PSB) adjusted posterior model-averaged effect size assuming it is present (i.e., without averaging over the point null models to reduce shrinkage toward zero), $\mu_{\text{adj},k}$; publication bias adjusted posterior probability of the presence of the effect, $p_{\text{adj},k}(\mathcal{H}_1 \mid \text{data}_k)$; and the posterior probability of the presence of PSB, $p_{\text{adj},k}(\mathcal{H}_\text{psb} \mid \text{data}_k)$. To isolate the effect of PSB adjustment, we compare the Bayesian, PSB unadjusted, model-averaged meta-analysis by dropping the PSB adjustment and thereby estimating the unadjusted posterior probability of the presence of the effect assuming it is present, $p_{\text{unadj},k}(\mathcal{H}_1 \mid \text{data}_k)$.\end{revision} $k = 1, \dots, K$ to denotes the individual meta-analyses. Each meta-analysis is based on $N_k$ estimates that are characterized with data describing the effect size $y_{k,n}$ and standard error $\text{se}_{k,n}$.

\subsubsection{Evidence for the Effect}

\begin{revision}We used the change in the posterior probability of the effect and the (standardized) evidence inflation factor to quantify the effect of PSB on meta-analytic evidence.\end{revision}

The \textbf{Posterior probability of the effect} \begin{revision}is an intuitive way of quantifying the evidence in favor of the alternative hypothesis of the presence of an effect. Under the assumption of equal prior probability of the presence and the absence of the effect, $p(\mathcal{H}_1) = p(\mathcal{H}_0)$, posterior probabilities larger than 0.5 indicate that the data are more likely under the presence of the effect. On the other hand, posterior probabilities lower than 0.5 indicate that the data are more likely under the absence of the effect. The ability to quantify evidence for both the null and the alternative is a key benefit of Bayesian methods over null hypothesis significance testing. \cite{wagenmakers2016bayesian, wasserstein2016asa}.\end{revision}

\begin{revision}A corresponding way of quantifying the evidence of an effect is via Bayes factors (see InfoBox~\ref{box:Bayesfactor} for more detail). Bayes factors quantify the change from prior to posterior odds for the presence of the effect. The advantage of Bayes factors is that they are independent of the prior odds for the presence of the effect. In other words, Bayes factors isolate the evidence for the presence of the effect contained in the data. In our settings, the assumption of equal prior probabilities leads to an equivalence between Bayes factors and posterior odds.\end{revision}

\begin{revision}The change from the PSB unadjusted posterior probability of the effect, $p_{\text{unadj},k}(\mathcal{H}_1 \mid \text{data}_k)$, to the PSB adjusted posterior probability of the effect, $p_{\text{adj},k}(\mathcal{H}_1 \mid \text{data}_k)$, quantifies the amount of evidence introduced by PSB. The larger the impact of PSB, the larger the difference between the PSB unadjusted and PSB adjusted posterior probabilities of the effect. If there was no PSB, we would observe no change in the posterior probability of the effect after PSB adjustment.\end{revision}

\textbf{Evidence inflation factor} \begin{revision}(EIF) quantifies the degree to which the evidence in favor of the presence of the effect was inflated due to PSB. $\text{EIF}_k$ quantifies the amount of evidence in favor of the effect in the PSB unadjusted meta-analysis, $\text{BF}_{10, \text{unadj},k}$, to the amount of evidence in favor of the effect in the PSB adjusted meta-analysis $\text{BF}_{10, \text{adj},k}$,\end{revision}
\begin{equation}
    \text{EIF}_k = \frac{\text{BF}_{10, \text{unadj},k}}{\text{BF}_{10, \text{adj},k}}.
\end{equation}
\noindent An evidence inflation factor larger than one indicates inflated evidence in favor of the effect due to PSB. 

\begin{revision}However, the amount of evidence contained in each meta-analysis, and the corresponding evidence inflation, is dependent on the number of meta-analyzed estimates, $N_n$, i.e., more estimates leads to more evidence. To facilitate the comparison of evidence inflation due to PSB in meta-analyses with different numbers of estimates, we also compute the standardized, per-estimate, evidence inflation factor in each meta-analysis, $\text{sEIF}_k$,\end{revision} by standardizing the EIF by the number of estimates,
\begin{equation}
    \text{sEIF}_k = \text{EIF}_k^{\frac{1}{N_k}},
\end{equation}
\noindent \begin{revision}where sEIF represents each estimate's marginal contribution, on average, to the evidence inflation due to PSB. The sEIF also partially mitigates the potential issue of dependent estimates within a meta-analysis. In the most extreme case, e.g., identical estimates, the same data is conditioned upon multiple times, which leads to overestimation of evidence. Taking only a fraction of each estimate's likelihood, proportional to the number of estimates, then ensures that the data are not conditioned upon more than once, although data producing multiple estimates are still weighted more heavily.\end{revision}

\subsubsection{Effect Size Estimates}

\textbf{Absolute bias} (bias) quantifies the degree to which the average effect sizes in each meta-analysis,
\begin{equation*}
    \Hat{y}_k = \frac{1}{N_k} \sum_{n = 1}^{N_k} y_{k,n},
\end{equation*}
overestimates the PSB-adjusted meta-analytic effect size estimate assuming the presence of the effect $\mu_{\text{adj},k}$, 
\begin{equation}
    \text{bias}_k = \Hat{y}_k - \mu_{\text{adj},k}.
\end{equation}
\noindent Absolute bias larger than zero indicates that PSB leads to inflated effect size estimates. We compare the average effect sizes to the PSB-adjusted effect sizes assuming the presence of the effect (conditional effect size estimates) rather than averaging across all models. Excluding models assuming the absence of a mean effect mitigates the pooling towards $0$ in meta-analyses more consistent with the null hypothesis. Tables 5 and 6 in the Supplementary Materials use the PSB-adjusted effect sizes model-averaged across all models, including models assuming the absence of a mean effect. These estimates, which are model-averaged also over the null, indicate stronger absolute bias compared to the conditional estimates presented in the main manuscript (Table 2).

\textbf{Overestimation factor} (OF) quantifies the degree to which the average effect sizes in meta-analyses overestimate the PSB-adjusted  effect size estimates assuming the presence of the effect,
\begin{equation}
    \text{OF} = \frac{ \frac{1}{K}\sum_{k=1}^{K} \Hat{y}_k  }{ \frac{1}{K}\sum_{k=1}^{K}  \mu_{\text{adj},k} }.
\end{equation}
\noindent An overestimation factor larger than one is evidence of PSB. We use delta method to obtain confidence interval of the overestimation factor. In the Supplementary Materials, we also report medians and interquartile ranges of per meta-analysis overestimation factors,
\begin{equation}
    \text{OF}_k = \frac{\Hat{y}_k}{\mu_{\text{adj},k}}.
\end{equation}
\noindent However, note that $\text{OF}_k$ can lead to non-sensible results as a meta-analysis with a positive mean effect and very small negative PSB-adjusted  effect sizes estimate results in an extremely large negative $\text{OF}_k$.

\subsubsection{Evidence for Publication Selection Bias}

\textbf{Posterior probability of PSB} \begin{revision}is an intuitive way of quantifying the evidence in favor of PSB. Similarly to the posterior probability of the presence of the effect, under the assumption of equal prior probability of the presence and the absence of PSB, $p(\mathcal{H}_\text{PSB}) = p(\mathcal{H}_{\text{NoPSB}})$, a posterior probability larger than 0.5 indicates that the data are more likely under the presence of PSB. On the other hand, a posterior probability lower than 0.5 indicates that the data are more likely under the absence of PSB. As before, the Bayes factor for the presence of PSB, $\text{BF}_{\text{psb}}$,\footnote{In other publications, we abbreviate this as $\text{BF}_{\text{pb}}$ or $\text{BF}_{\omega\Bar{\omega}}.$} quantifies the change from prior to posterior odds for the presence of PSB. A Bayes factor in favor of the presence of PSB larger than one provides evidence in favor of the presence of PSB and lower than one provides evidence against the presence of PSB\end{revision}

\textbf{Relative publication probabilities} quantify the relative probability of an estimate being published for a given $p$-value interval compared to estimates with statistically significant $p$-values. We use one-sided $p$-values, resulting in $p$-values larger than 0.5 corresponding to estimates in the opposite direction. To facilitate the interpretation we visualize a weight function that shows the change of relative publication probabilities across the range of $p$-values. We report the results only in Supplementary Materials.

\textbf{Effect size inflation in imprecise estimates} quantifies the relationship between the effect sizes and their standard errors. To facilitate the interpretation of the funnel asymmetry test, we visualize the bias in effect sizes as a function of standard errors (incorporating the quadratic term from the RoBMA model). We report the results only in Supplementary Materials.

\section{Results}

\subsection{Descriptives}

\begin{table}[h]
    \caption{Summary of the data sets from each field. The number of estimates per meta-analysis (Estimates/MA) and the \begin{revision}unweighted simple \end{revision}mean effect size of estimates within each meta-analysis (Effect Sizes) are reported as medians with the interquartile range \begin{revision}(in parentheses)\end{revision}. The proportion of the statistically significant (Prop. Significant) \begin{revision}meta-analytic effect size estimates\end{revision} is based on a random-effect meta-analysis estimated via restricted maximum likelihood with $\alpha = 0.05$ (removing one environmental sciences and 275 medical meta-analyses that did not converge) .}
    \begin{tabular}{rrrrrc}
    \hline
    Field         & Meta-Analyses & Estimates & Estimates/MA & Effect Sizes ($d$) & Prop. Significant \\ 
    \hline
    Medicine      & 67,386 & 597,699 &    5 (4, 10) & 0.24 (0.09, 0.47) & 0.39 \\ 
    Environmental &    199 &  12,707 &  26 (11, 59) & 0.62 (0.31, 0.95) & 0.85 \\ 
    Psychology    &    605 &  23,563 &   18 (9, 40) & 0.37 (0.18, 0.61) & 0.78 \\ 
    Economics     &    327 &  91,421 & 66 (30, 283) & 0.20 (0.09, 0.37) & 0.82 \\
    \hline
    \end{tabular}
    \label{tab:descriptives}
\end{table}

\begin{revision}Table~\ref{tab:descriptives} compares the characteristics of the meta-analyses from each field. Medical meta-analyses contain the smallest number of estimates per meta-analysis, followed by psychology and environmental sciences with five to six times the number of estimates compared to medicine. Finally economics meta-analyses contain over twelve times the median number of estimates compared to medicine. Contrary to a naive expectation that more estimates may be conducted to establish smaller effects, economic and medical effect sizes are approximately the same magnitude (measured as a mean effect size per meta-analysis). Effect sizes in psychology are roughly twice as large as those in economics, and effect sizes in the environmental sciences are approximately three times larger than those in economics. Differences in the number of estimates per meta-analysis are most closely reflected in the proportion of statistically significant random-effects estimates. Notably, random-effects estimates in economics, psychology, and environmental sciences are statistically significant approximately twice as often as in medicine. This disparity in the proportion of statistically significant meta-analyses is consistent when comparing meta-analyses with a matched number of estimates across the disciplines, although the difference in mean effects is somewhat smaller (see Table 1 in the Supplementary Materials).\end{revision}

We summarize results from all meta-analyses, apart from seven medical meta-analyses that did not converge. See Supplementary Materials for analyses showing that matching meta-analyses based on the number of primary estimates within each meta-analysis does not meaningfully affect the conclusions.

\subsection{Evidence for the Effect}

\begin{figure}[h]
    \caption{\begin{revision}Median, Interquartile Range, and\end{revision} Distribution of Posterior Probability for the Presence of the Effect Before and After Adjustment for Publication Selection Bias in Each Field}
    \includegraphics[width=6.5in]{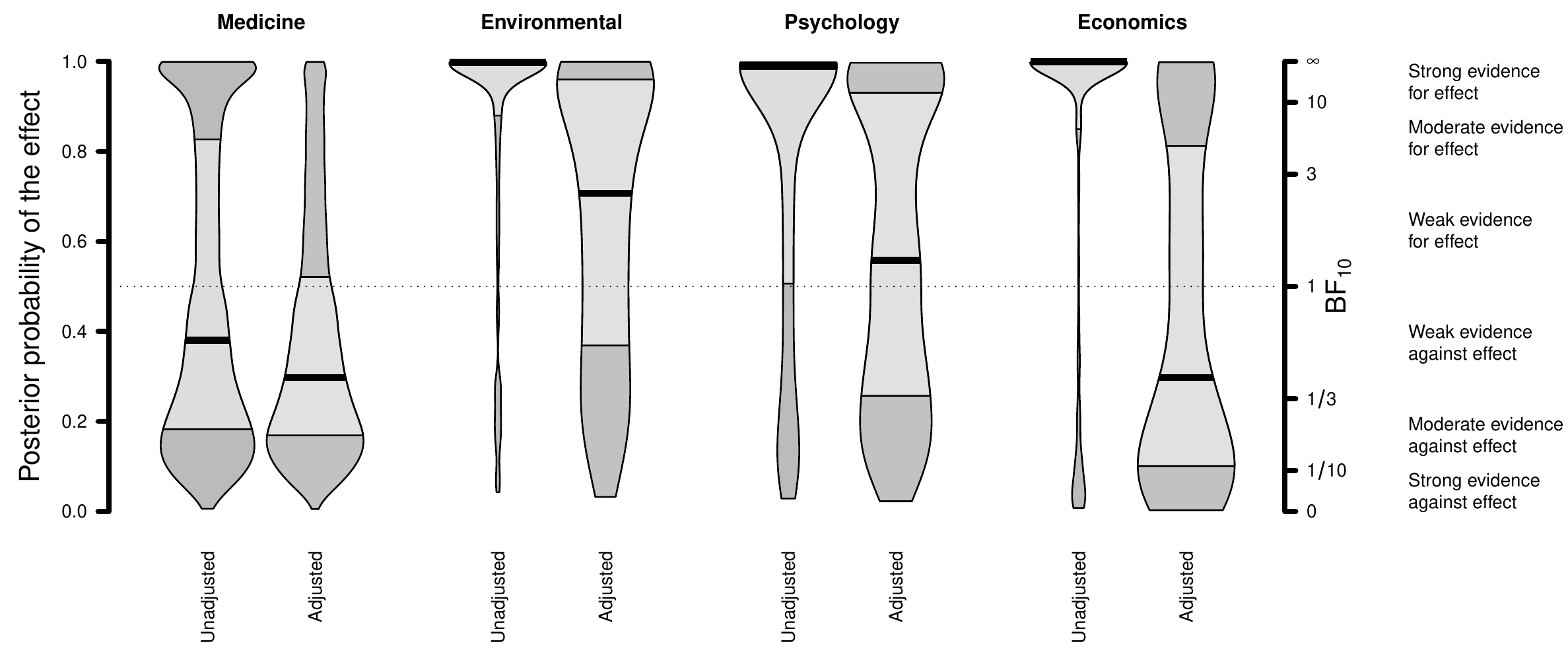}
    \newline
    \label{fig:post_prob}
    \textit{Note.} The width of grey area indicates density, the light grey area indicates the interquartile range, and the black line indicates the median. The $y$-axis is scaled according to posterior probabilities assuming equal prior probabilities of presence vs absence of the effect. See the secondary $y$-axis for Bayes factors in favor of the effect that are independent of the assumed prior probability of the effect.
\end{figure}

Figure~\ref{fig:post_prob} shows \begin{revision}medians, interquartile ranges, and\end{revision} distributions of the posterior probability of an effect before and after adjusting for PSB. These distributions reveal several patterns. First, meta-analyses in economics, psychology, and environmental sciences predominantly show evidence for an effect before adjusting for PSB (unadjusted); whereas meta-analyses in medicine often display evidence against an effect. This disparity between the fields remains even when comparing meta-analyses with equal numbers of effect size estimates (see Supplementary Materials). After correcting for PSB, the posterior probability of an effect drops much more in economics, psychology, and environmental sciences (medians drop from $99.9$\% to $29.7$\%, from $98.9$\% to $55.7$\%, and from $99.8$\% to $70.7$\% , respectively) compared to medicine ($38.0$\% to $29.7$\%). The pattern is especially striking in economics, where the median posterior probability of an effect drops by more than seventy percentage points after PSB correction. Mean decreases in posterior probabilities show a similar pattern but with somewhat smaller reductions (Table 9 and 10 in Supplementary Materials). \begin{revision} In all four disciplines, adjusting for PSB resulted in a substantial decrease in the strength of evidence for the effect: the proportion of meta-analyses with at least strong evidence for the presence of an effect (i.e., $\text{BF}_{10} > 10$) decreased from 20.2\% to 5.3\% in medicine, from 72.4\% to 30.7\% in environmental sciences, from 59.8\% to 27.3\% in psychology, and from 72.8\% to 19.6\% in economics. A comparable decrease was also present when comparing the proportion of meta-analyses with at least moderate evidence for the presence of an effect (i.e., $\text{BF}_{10} > 3$; from 28.9\% to 12.3\% in medicine, from 80.4\% to 47.3\% in environmental sciences, from 67.4\% to 38.39\% in psychology, and from 76.8\% to 27.6\% in economics).\end{revision}

%
%
%

Furthermore, we quantify the inflation of evidence in favor of an effect in meta-analyses via the evidence inflation factor -- the increase in Bayes factor in favor of the effect due to the PSB. We find that meta-analyses in economics inflate the evidence by a median factor of 11,369, whereas the meta-analyses in environmental sciences and psychology inflate the evidence by `only' $45.9$ and $30.0$ respectively, and medicine by a median factor of $1.33$. These extreme differences between the fields are largely driven by the disparity in the typical numbers of estimates per meta-analysis across the disciplines (Table 1). \begin{revision}After standardizing the evidence inflation factor (sEIF) by the number of estimates per meta-analysis, we find\end{revision} that per estimate evidence inflation is the largest in psychology with a median factor of $1.27$, followed by environmental sciences with a median factor of $1.22$, economics with a median factor of $1.15$, and medicine with a median factor of $1.05$. Again, the strong evidence inflation in economics (11,369) is largely due to having many more estimates per meta-analysis than psychology and medicine. \begin{revision}However, even after adjusting for different numbers, meta-analyses in medicine still show the least inflated evidence due to PSB.\end{revision}

\subsection{Effect Size Estimates}

\begin{table}[h]
    \caption{Summary of the footprints of publication selection bias on the meta-analytic effect sizes in the form of absolute bias (in Cohen's $d$) and overestimation factor.}
    \begin{tabular}{rrrr}
    Field         & Absolute Bias ($d$) & Overestimation Factor \\ 
    \hline
    Medicine      & 0.13 [0.12, 0.13]  & 1.62 [1.60, 1.64] \\ 
    Environmental & 0.33 [0.24, 0.41]  & 1.78 [1.42, 2.13] \\ 
    Psychology    & 0.13 [0.11, 0.14]  & 1.39 [1.24, 1.55] \\ 
    Economics     & 0.15 [0.13, 0.17]  & 2.16 [1.69, 2.64] \\
    \hline
    \end{tabular}
    \\
    \textit{Note.} The results are based on the comparison of publication bias adjusted meta-analytic effect size estimates assuming presence of the effect to the mean effect sizes per meta-analysis. The table displays means and 95\% confidence intervals. (See Table 4 in the Supplementary Materials for medians and interquartile ranges.)
    \label{tab:es_measures_mean}
\end{table}

\begin{revision}Table~\ref{tab:es_measures_mean} summarizes the effect of PSB on effect sizes in each field. The first column reveals that environmental sciences, on average, suffer from as much as two and a half times larger absolute bias as medicine, economics, or psychology. The degree of absolute bias in environmental sciences is so large that it is comparable to average unadjusted effect sizes in other fields. Otherwise, medicine, psychology, and economics share a comparable degree of absolute bias. The median absolute bias in each field is lower than the mean bias due to the right skew distribution of absolute biases (see Table 4 in the Supplementary Materials). \end{revision}

\begin{revision}The second column of Table~\ref{tab:es_measures_mean} displays the relative impact of PSB on meta-analytic estimate via the overestimation factor. On average, economics meta-analyses are, relatively, the most PSB exaggerated, inflating effect sizes by over two times; this corroborates a prior survey on power and bias \cite{ioannidis2017power}.  Effect sizes in environmental sciences and medical meta-analyses show smaller yet notable relative effect size inflation. Finally, effect sizes in psychological meta-analyses are the least inflated with the average effect size exaggerated by 40\%. In each field, the distribution of absolute biases is right-skewed; consequently, the per meta-analysis overestimation factor \emph{median} is lower than the mean (see Table 4 in the Supplementary Materials). The median overestimation factor is relatively stable/decreasing with the increasing number of effect size estimates per meta-analysis, suggesting that the number of meta-analyses does not play a role in the relative size of PSB.\end{revision}

\subsection{Evidence for Publication Selection Bias}

\begin{figure}[h]
    \caption{\begin{revision}Median, Interquartile Range, and\end{revision} Distribution of Posterior Probability for the Presence of Publication Selection Bias in Each Field}
    \includegraphics[width=5in]{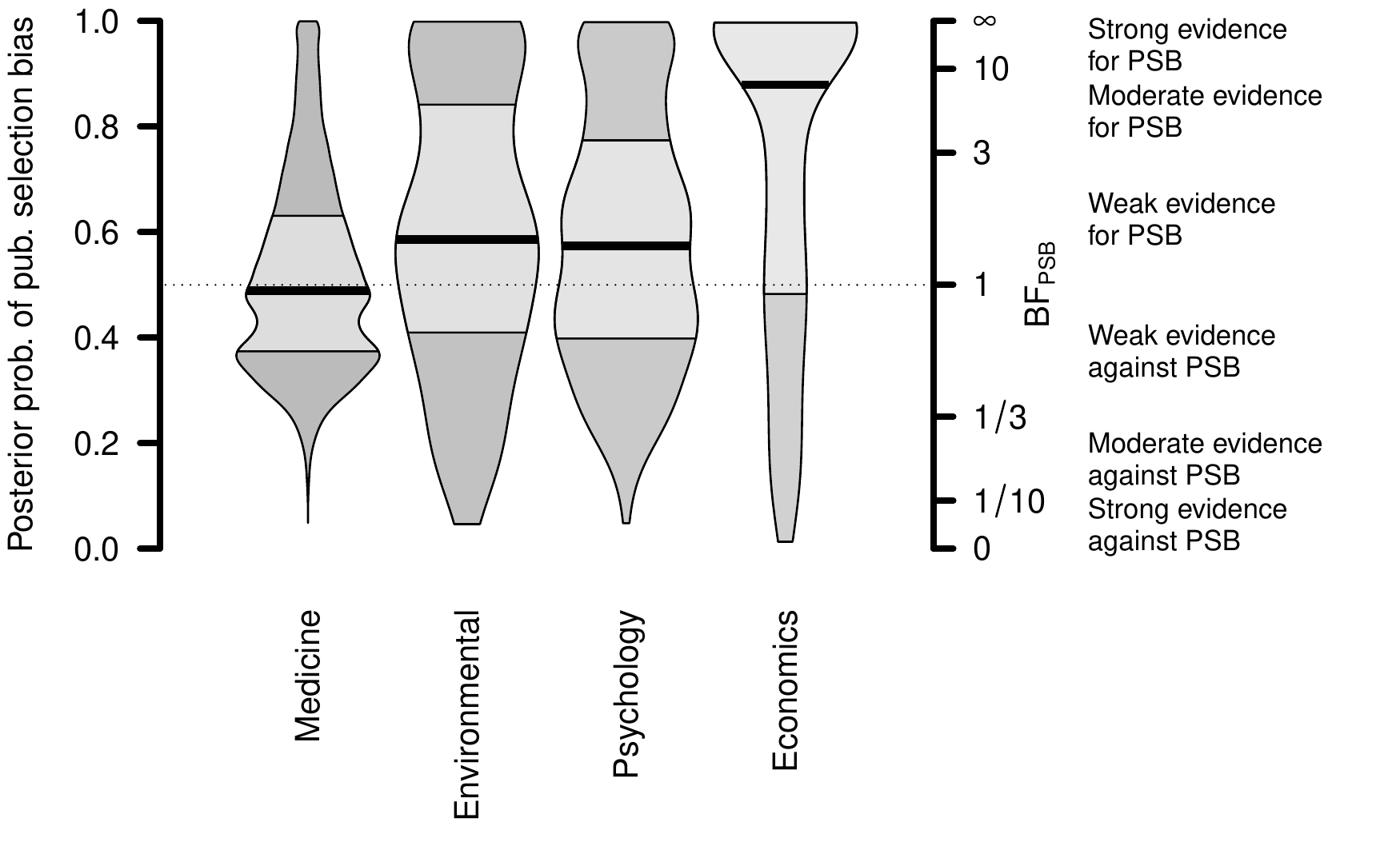}
    \newline
    \label{fig:pub_bias}
    \textit{Note.} The width of grey area indicates density, the light grey area indicates the interquartile range, and the black line indicates the median. The $y$-axis is scaled according to posterior probabilities, assuming equal prior probabilities of presence vs. absence of the publication selection bias. See the secondary $y$-axis for Bayes factors in favor of the publication selection bias that are independent of the assumed prior probability of the publication selection bias.
\end{figure}

\begin{revision}Figure~\ref{fig:pub_bias} shows medians, interquartile ranges, and distributions of the posterior probability of the PSB in each field. We find the most evidence for PSB in economics, where the typical evidence of the presence of publication bias is moderate, median $\text{BF}_\text{PSB}=7.27$ corresponding to $87.9$\% posterior probability of publication selection bias. Meta-analyses in environmental sciences and psychology have weak evidence in favor of PSB; even though the proportion of meta-analyses showing at least moderate evidence for PSB is still considerable (32.2\% and 27.4\% respectively). Meta-analyses in medicine show the lowest proportion of at least moderate evidence in favor of PSB (12.9\%). However, the proportion of meta-analyses consistent with the evidence of absence of PSB (i.e., $\text{BF}_{\text{PSB}} < 1/3$) is also the lowest in medicine (2.6\%), indicating that the majority of medical meta-analyses is not informative enough to provide compelling evidence for or against publication bias. The proportion of meta-analyses with at least moderate evidence against PSB is somewhat higher in psychology (7.1\%), economics (12.2\%), and environmental sciences (12.6\%). Meta-analyses with a larger number of effect size estimates present slightly more evidence in favor of the PSB; however, the overall disparity between the fields remains when comparing meta-analyses with a matched number of effect size estimates (Table 17 in Supplementary Materials). \end{revision}

\section{Concluding Comments}

We present a comprehensive assessment of publication selection bias and its effects on meta-analyses across medicine, environmental sciences, psychology, and economics. Novel methods and measures allowed us to quantify the evidence for the absence or presence of the mean effect and publication selection bias, as well as inflation of the evidence of the effect due to the publication selection bias. Furthermore, we estimated the bias and overestimation factor of the effect sizes of average estimates included in meta-analyses.

Our analysis is based on all effect size estimates found in these meta-analyses, regardless of the type outcome or how they were analyzed. One can classify outcomes into three categories. First, some outcomes may have been pre-specified as being of primary interest to show a desirable effect (e.g., the effectiveness of a medication in reducing the risk of death). Second, some other outcomes are not pre-specified but may still be used to demonstrate some preferred outcome; thus, they may have larger analytical flexibility (e.g., using alternative measures of effectiveness) and thereby are potentially more affected by publication selection bias. Third, still other outcomes may have been collected and analyzed without any strong interest to show some significant result, or even with some incentive to show non-significant results (e.g., outcomes on collected adverse events). Publication selection bias is expected mostly in the second category, while it may be less in the first category \cite{ioannidis1998issues} and may be entirely absent in the third category.

\begin{revision}Furthermore, we assumed independence of the reported primary estimates within and between meta-analyses; that is, each reported estimate is regarded to provide the same amount of new information as every other reported estimate. However, estimates may be dependent between meta-analyses, e.g., a single estimate might be used across multiple meta-analyses, and within meta-analyses, e.g., multiple estimates obtained from a single study/primary data set. As our data does not allow us to tackle those dependencies directly, we discuss how each independence violation might affect the results. The between meta-analysis dependency of estimates is of lesser importance as our inferences are concerned with the population of meta-analyses. Consequently, between meta-analysis dependency of estimates would only affect descriptive summaries of the estimates themselves. The within meta-analysis dependency of estimates is more problematic and can lead to 1) the overestimation of the strength of evidence, as the same primary data set is conditioned upon more than once, and 2) placing more weight on studies with multiple estimates. The first issue is partially mitigated via the standardized evidence inflation factor, which assesses the average evidence contribution of an estimate, i.e., adjusting for the number of data sets conditioned upon. However, the absolute measures of evidence (i.e., evidence for the presence of the effect before and after publication bias adjustment or the evidence for publication bias) can be susceptible to overestimation, particularly in fields with relatively large within meta-analysis dependencies such as economics or environmental science (but see Supplementary Materials for comparison of meta-analyses with a matched number of effect size estimates). The second issue cannot be directly addressed; however, all presented measures are based on comparisons of two sets of models, both of which should be affected to a similar degree, thus hopefully canceling most of the bias that is generated by overweighting studies with multiple estimates. Overall, we cannot exclude that the observed between-field differences may at least partially result from systematic differences in how meta-analyses themselves are conducted.\end{revision}

The milder publication selection bias in medical meta-analyses corroborates previous findings and might have multiple concurring explanations \cite{fanelli2017meta, fanelli2010positive, sterling1959publication}. \begin{revision} First, as in other disciplines, a large share of those medical meta-analyses with seemingly strong evidence no longer had strong evidence when PSB adjustment was made. However, a much lower proportion of medical meta-analyses showed strong evidence of an effect compared to the other disciplines. Therefore, the difference between medicine and the other disciplines might be explained by the higher proportion of meta-analyses in medicine that showed weak evidence for an effect already before adjusting for publication selection bias.\end{revision} Second, medical studies may measure phenomena that are simpler and more stable, using methods that are more solidly and universally codified, which reduces researchers' ``degrees of freedom'' in generating and publishing evidence \cite{fanelli2017meta, fanelli2010positive}. Third, it is also possible that the milder publication selection bias seen in medical meta-analyses is reflecting a larger share of meta-analyses that belonged to a category of outcomes with less pressure for publication selection bias. Finally, medical research makes wider use of research integrity practices, such as clinical trial registration, which might reduce the risk of publication selection bias \cite{laine2007clinical}. Perhaps, medical research is, therefore, typically of a higher methodological quality and less subject to bias \cite{fanelli2017meta}.

In this paper, we documented the considerable impact of publication selection bias on meta-analyses in a variety of disciplines. Even though we can probe the footprint of these biases with the statistical techniques employed here, science ultimately needs to progress toward mitigating publication bias already while conducting and publishing the research. While the specific patterns of researchers' ``degrees of freedom'' and causes of publication selection bias are likely to vary widely across fields, our results suggest that the social sciences might especially benefit from adopting practices to mitigate these, including: preregistration, greater transparency, and registered reports \cite{chambers2013registered, chambers2015registered, vanakker2021preregistration}.

\section*{Highlights}
- Publication selection bias, where studies with significant or positive results are more likely to be reported and published, distorts the available scientific record.\\
- This study surveyed over 68,000 meta-analyses from medicine, environmental sciences, psychology, and economics to assess the extent of publication selection bias. As a result, it underscores the importance of addressing publication bias in evidence synthesis\\
- Results suggest that meta-analyses in economics are the most affected by publication selection bias, followed by environmental sciences and psychology. In contrast, meta-analyses in medicine are suggested to be the least affected. Yet, notable biases are found across all of these scientific disciplines.\\
- This study documents the potential extent of publication bias in different fields, which could help researchers and the public better understand the limitations of research and the potential biases of research synthesis.\\

\section*{Availability of Data and Materials}
See \url{https://osf.io/bgfzp/} for data and analysis scripts.

\bibliographystyle{WileyNJD-AMA}
\bibliography{manuscript.bib}

\end{document}


\maketitle 

\newpage

\section{Additional Results}
\label{app:results2}

\begin{table}[H]
\caption{Proportion of statistically significant meta-analytic effect size estimates based on random effects meta-analysis estimated via restricted maximum likelihood for meta-analyses with different number of studies.}
  \footnotesize	
  \begin{tabular}{rrrrrrrrr}
  \hline
  & \multicolumn{2}{c}{Medicine} & \multicolumn{2}{c}{Economics} & \multicolumn{2}{c}{Psychology} & \multicolumn{2}{c}{Environmental} \\
  K & Prop. Significant & K & Prop. Significant & K & Prop. Significant & K & Prop. Significant & K \\ 
  \hline
      $<$ 5 & 0.29 & 26885 &  &   0 & 0.33 &  55 & 1.00 &   3 \\ 
     5 - 9 & 0.39 & 23316 & 0.58 &  12 & 0.66 & 114 & 0.86 &  42 \\ 
    10 - 19 & 0.52 & 11554 & 0.73 &  33 & 0.84 & 159 & 0.90 &  42 \\ 
    20 - 29 & 0.61 & 2821 & 0.86 &  36 & 0.85 &  79 & 0.62 &  16 \\ 
    30 - 49 & 0.62 & 1679 & 0.81 &  58 & 0.84 &  74 & 0.81 &  37 \\ 
   50 - 99 & 0.79 & 705 & 0.77 &  52 & 0.88 &  68 & 0.90 &  31 \\ 
  100 - 299 & 0.60 & 141 & 0.83 &  59 & 0.92 &  51 & 0.82 &  17 \\ 
    $\geq$ 300 & 1.00 &   3 & 0.92 &  77 & 1.00 &   5 & 0.90 &  10 \\ 
  \hline
\end{tabular}
    \label{tab:prop_significant_k}
\end{table}

\begin{table}[h]
    \caption{Means and 95\% confidence intervals of the posterior probability of an effect before and after adjusting for PSB, and the posterior probability of PSB.}
    \begin{tabular}{rrrrr}
    Field         & Effect (unadjusted) & Effect (adjusted) & PSB \\ 
    \hline
    Medicine      &  0.48 [0.48, 0.48] & 0.37 [0.37, 0.38] & 0.52 [0.52, 0.52] \\ 
    Economics     &  0.80 [0.75, 0.84] & 0.44 [0.38, 0.49] & 0.59 [0.55, 0.63] \\ 
    Psychology    &  0.76 [0.73, 0.79] & 0.57 [0.53, 0.61] & 0.73 [0.68, 0.78] \\ 
    Environmental &  0.85 [0.80, 0.90] & 0.65 [0.58, 0.71] & 0.60 [0.53, 0.67] \\
    \hline
    \end{tabular}
\end{table}

\begin{table}[h]
    \caption{Means and 95\% confidence intervals for the evidence inflation factors (EIF) and standardized evidence inflation factors (sEIF) in each field. Two economics and three medical meta-analysis were omited from the computation as the evidence inflation factor was larger than the numerical precision of R.}
    \begin{tabular}{rrr}
    Field         & EIF & sEIF \\ 
    \hline
    Medicine      & 3.62 [3.53, 3.72]                                                & 1.15 [1.15, 1.15] \\
    Economics     & 1.9 $\times 10^{10}$ [1.8 $\times 10^{8}$, 2.1 $\times 10^{12}$] & 1.20 [1.17, 1.22] \\
    Psychology    & 1.9 $\times 10^{5}$  [2.7 $\times 10^{4}$, 1.3 $\times 10^{6}$]  & 1.32 [1.29, 1.34] \\
    Environmental & 4.2 $\times 10^{3}$  [0.8 $\times 10^{3}$, 2.0 $\times 10^{5}$]  & 1.28 [1.24, 1.32] \\
    \hline
    \end{tabular}
\end{table}

\begin{table}[H]
    \caption{Medians and interquartile ranges of absolute bias (in Cohen's $d$) and per meta-analysis overestimation factor.}
    \begin{tabular}{rrrr}
    Field      & Absolute bias ($d$) & Overestimation Factor \\ 
    \hline
    Medicine      & 0.09 (0.01, 0.20) & 1.26 (0.60, 1.96)  \\ 
    Economics     & 0.07 (0.02, 0.20) & 1.18 (0.73, 2.31)  \\ 
    Psychology    & 0.07 (0.02, 0.16) & 1.15 (1.02, 1.50)  \\ 
    Environmental & 0.14 (0.04, 0.34) & 1.18 (1.01, 1.56)  \\
    \hline
    \end{tabular}
    \label{tab:es_measures_median}
\end{table}

\section{Additional Measures of Publication Selection Bias}

\begin{figure}[H]
    \caption{Distribution of posterior probability for the publication selection bias (A), different types of the publication bias assuming it is present (B), the relative publication probabilities for individual studies (C), and effect size inflation of individual studies as a function of sample size (D) for each field.}
    \includegraphics[width=5in]{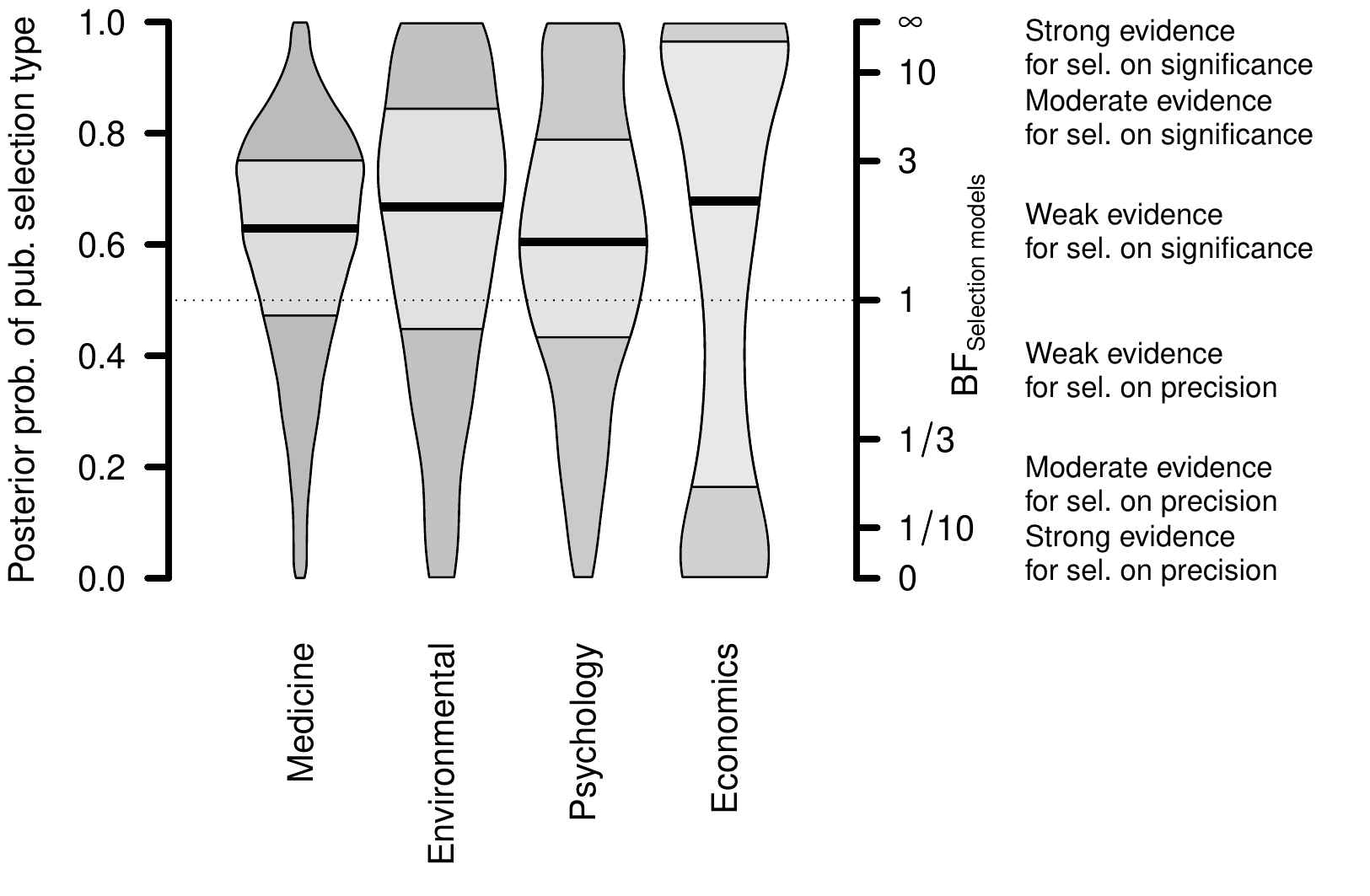}
    \newline
    \textit{Note.} The width of grey area indicates density, the light grey area indicates the interquartile range, and the black line indicates the median. The $y$-axe is scaled according to posterior probabilities assuming equal prior probabilities of presence vs. absence of the type of publication selection bias. See the secondary $y$-axis for Bayes factors in favor of the effect that are not dependent on the assumed prior probabilities. (We display only a random sample of 5,000 meta-analyses from medicine to enhance clarity.)
    \label{fig:bias_type}
\end{figure}

Figure~\ref{fig:bias_type}  presents results regarding the mode in which PSB operates. Most meta-analyses provide slightly more evidence in favor of selection models compared to effect size inflation due to small studies, with the posterior probability of selection models assuming the presence of PSB in medicine (median/mean = $62.9/59.8$\%), economics (median/mean = $67.9/56.9$\%), psychology (median/mean = $60.5/59.7$\%), and environmental sciences (median/mean = $66.8/62.3$\%).

\begin{figure}[H]
    \caption{The relative publication probabilities for individual studies (A), and effect size inflation of individual studies as a function of sample size (B) for each field.}
    \includegraphics[width=6.5in]{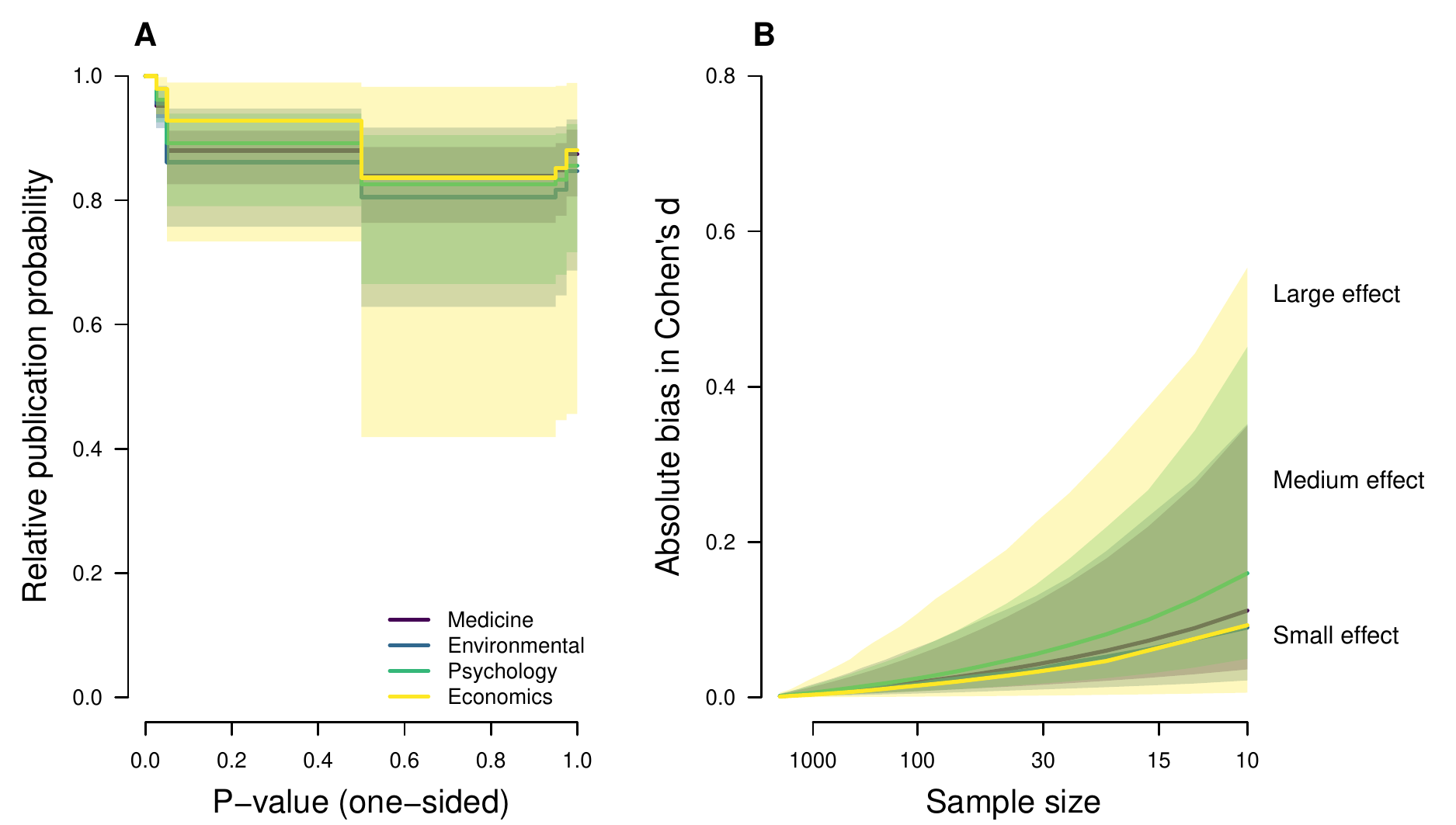}
    \newline
    \textit{Note.} Full lines correspond to medians with the shaded areas depicting interquartile ranges.
    \label{fig:bias_type2}
\end{figure}

Figure~\ref{fig:bias_type2} shows how the PSB affects the probability of the individual studies being published based on their $p$-values (A) and the expected effect size inflation in small studies (B). Visualization of the median and interquartile range of mean relative publication probabilities in each field (i.e., the relative probability of non-significant $p$-values to be published in comparison to significant $p$-values) shows us comparable median relative publication probabilities for all fields, however, we find lower relative publication probabilities for the lower quartiles in psychology and especially economics. Visualization of the median and interquartile range of effect size inflation as a function of sample size of the primary studies in each field provides sligtly different results. We see a slightly steeper increase in bias for very small studies in environmental sciences and psychology in comparison to  economics and medicine, however, the upper quartiles of the bias is, again, larger in economics than psychology, medicine or environmental sciences.

\section{Additional Results of Effect Sizes}

\begin{table}[H]
    \caption{Means and 95\% CI of absolute bias (in Cohen's $d$) and overestimation factor model-averaged across models assuming absence of the effect.}
    \begin{tabular}{rrrr}
    Field         & Absolute bias ($d$) & Overestimation Factor \\ 
    \hline
    Medicine      & 0.20 [0.20, 0.20]  & 2.54 [2.50, 2.58] \\ 
    Economics     & 0.18 [0.16, 0.20]  & 2.88 [2.16, 3.59] \\ 
    Psychology    & 0.18 [0.16, 0.20]  & 1.68 [1.47, 1.89] \\ 
    Environmental & 0.40 [0.31, 0.48]  & 2.16 [1.69, 2.62] \\
    \hline
    \end{tabular}
    \label{tab:es_measures_pooled}
\end{table}

\begin{table}[H]
    \caption{Medians and interquartile ranges of absolute bias (in Cohen's $d$) and per meta-analysis overestimation factor model-averaged across models assuming absence of the effect.}
    \begin{tabular}{rrrr}
    Field         & Absolute bias ($d$) & Overestimation Factor \\ 
    \hline
    Medicine      & 0.15 (0.06, 0.29) & 2.71 (1.08, 7.75)  \\ 
    Economics     & 0.12 (0.04, 0.25) & 1.93 (0.91, 11.84) \\ 
    Psychology    & 0.13 (0.06, 0.24) & 1.56 (1.07, 4.37)  \\ 
    Environmental & 0.21 (0.07, 0.48) & 1.51 (1.06, 3.32)  \\
    \hline
    \end{tabular}
    \label{tab:es_measures_median_pooled}
\end{table}
\newpage

\section{Matched K Analysis}

\begin{table}[H]
\caption{Median and interquartile range of mean effect size for meta-analyses with different number of studies (in Cohen's $d$).}
\footnotesize	
\begin{tabular}{rrrrrrrrr}
  \hline
  & \multicolumn{2}{c}{Medicine} &  \multicolumn{2}{c}{Economics} &  \multicolumn{2}{c}{Psychology} & \multicolumn{2}{c}{Environmental} \\
  K & Effect size ($d$) & K & Effect size ($d$) & K & Effect size ($d$) & K \\ 
  \hline
$<$ 5 & 0.25 (0.09, 0.50) & 26885 &  &   0 & 0.24 (0.11, 0.46) &  55 & 0.76 (0.76, 0.78) &   3 \\ 
  5 - 9 & 0.24 (0.09, 0.47) & 23378 & 0.35 (0.16, 0.51) &  12 & 0.28 (0.14, 0.47) & 114 & 0.71 (0.57, 1.00) &  42 \\ 
  10 - 19 & 0.22 (0.09, 0.43) & 11683 & 0.27 (0.11, 0.42) &  33 & 0.45 (0.26, 0.67) & 159 & 0.83 (0.50, 1.23) &  42 \\ 
  20 - 29 & 0.22 (0.09, 0.43) & 2862 & 0.28 (0.13, 0.45) &  36 & 0.42 (0.16, 0.58) &  79 & 0.58 (0.36, 0.98) &  16 \\ 
  30 - 49 & 0.21 (0.09, 0.41) & 1720 & 0.24 (0.11, 0.49) &  58 & 0.41 (0.22, 0.66) &  74 & 0.56 (0.24, 0.75) &  38 \\ 
  50 - 99 & 0.31 (0.16, 0.56) & 713 & 0.23 (0.07, 0.38) &  52 & 0.37 (0.16, 0.60) &  68 & 0.36 (0.20, 0.72) &  31 \\ 
  100 - 299 & 0.18 (0.13, 0.34) & 142 & 0.19 (0.08, 0.37) &  59 & 0.51 (0.24, 0.98) &  51 & 0.57 (0.19, 0.74) &  17 \\ 
  $\geq$ 300 & 0.58 (0.45, 0.65) &   3 & 0.13 (0.07, 0.22) &  77 & 0.33 (0.30, 0.35) &   5 & 0.28 (0.20, 0.95) &  10 \\ 
  \hline
\end{tabular}
\end{table}

\begin{table}[H]
\caption{Mean and 95\% CI of mean effect size for meta-analyses with different number of studies (in Cohen's $d$).}
\footnotesize	
\begin{tabular}{rrrrrrrrr}
  \hline
  & \multicolumn{2}{c}{Medicine} &  \multicolumn{2}{c}{Economics} &  \multicolumn{2}{c}{Psychology} & \multicolumn{2}{c}{Environmental} \\
  K & Effect size ($d$) & K & Effect size ($d$) & K & Effect size ($d$) & K \\ 
  \hline
$<$ 5 & 0.36 [0.35, 0.36] & 26885 &  &   0 & 0.33 [0.25, 0.41] &  55 & 0.77 [0.75, 0.80] &   3 \\ 
  5 - 9 & 0.33 [0.33, 0.34] & 23378 & 0.45 [0.20, 0.70] &  12 & 0.36 [0.30, 0.42] & 114 & 0.82 [0.66, 0.97] &  42 \\ 
  10 - 19 & 0.31 [0.30, 0.32] & 11683 & 0.33 [0.23, 0.43] &  33 & 0.49 [0.44, 0.54] & 159 & 0.97 [0.75, 1.18] &  42 \\ 
  20 - 29 & 0.30 [0.29, 0.32] & 2862 & 0.32 [0.25, 0.40] &  36 & 0.42 [0.35, 0.48] &  79 & 0.69 [0.42, 0.96] &  16 \\ 
  30 - 49 & 0.29 [0.28, 0.31] & 1720 & 0.33 [0.26, 0.40] &  58 & 0.46 [0.38, 0.54] &  74 & 0.60 [0.45, 0.75] &  38 \\ 
  50 - 99 & 0.38 [0.36, 0.40] & 713 & 0.29 [0.22, 0.36] &  52 & 0.44 [0.35, 0.53] &  68 & 0.53 [0.36, 0.70] &  31 \\ 
  100 - 299 & 0.23 [0.19, 0.27] & 142 & 0.28 [0.21, 0.35] &  59 & 0.79 [0.55, 1.02] &  51 & 0.72 [0.35, 1.09] &  17 \\ 
  $\geq$ 300 & 0.54 [0.30, 0.78] &   3 & 0.17 [0.13, 0.21] &  77 & 0.32 [0.29, 0.35] &   5 & 0.57 [0.21, 0.93] &  10 \\ 
  \hline
\end{tabular}
\end{table}

\begin{table}[H]
\caption{Median and interquartile range of posterior probability of the presence of the effect based on publication bias unadjusted Bayesian model-averaged meta-analysis for meta-analyses with different number of studies.}
\footnotesize	
\begin{tabular}{rrrrrrrrr}
  \hline
  & \multicolumn{2}{c}{Medicine} &  \multicolumn{2}{c}{Economics} &  \multicolumn{2}{c}{Psychology} & \multicolumn{2}{c}{Environmental} \\
  K & Post. prob. & K & Post. prob. & K & Post. prob. & K \\ 
  \hline
$<$ 5 & 0.35 (0.19, 0.65) & 26885 &  &   0 & 0.37 (0.23, 0.71) &  55 & 0.99 (0.93, 0.99) &   3 \\ 
  5 - 9 & 0.37 (0.17, 0.82) & 23378 & 0.60 (0.23, 0.96) &  12 & 0.75 (0.25, 0.95) & 114 & 0.99 (0.95, 1.00) &  42 \\ 
  10 - 19 & 0.47 (0.16, 0.96) & 11683 & 0.95 (0.24, 1.00) &  33 & 0.99 (0.80, 1.00) & 159 & 0.98 (0.88, 1.00) &  42 \\ 
  20 - 29 & 0.69 (0.16, 1.00) & 2862 & 0.99 (0.62, 1.00) &  36 & 1.00 (0.64, 1.00) &  79 & 0.72 (0.49, 1.00) &  16 \\ 
  30 - 49 & 0.76 (0.14, 1.00) & 1720 & 1.00 (0.65, 1.00) &  58 & 1.00 (0.98, 1.00) &  74 & 1.00 (0.92, 1.00) &  38 \\ 
  50 - 99 & 1.00 (0.44, 1.00) & 713 & 1.00 (0.59, 1.00) &  52 & 1.00 (0.93, 1.00) &  68 & 1.00 (0.90, 1.00) &  31 \\ 
  100 - 299 & 0.26 (0.08, 1.00) & 142 & 1.00 (1.00, 1.00) &  59 & 1.00 (1.00, 1.00) &  51 & 1.00 (0.94, 1.00) &  17 \\ 
  $\geq$ 300 & 1.00 (1.00, 1.00) &   3 & 1.00 (1.00, 1.00) &  77 & 1.00 (1.00, 1.00) &   5 & 1.00 (1.00, 1.00) &  10 \\ 
  \hline
\end{tabular}
\end{table}

\begin{table}[H]
\caption{Mean and 95\% CI of posterior probability of the presence of the effect based on publication bias unadjusted Bayesian model-averaged meta-analysis for meta-analyses with different number of studies.}
\footnotesize	
\begin{tabular}{rrrrrrrrr}
  \hline
  & \multicolumn{2}{c}{Medicine} &  \multicolumn{2}{c}{Economics} &  \multicolumn{2}{c}{Psychology} & \multicolumn{2}{c}{Environmental} \\
  K & Post. prob. & K & Post. prob. & K & Post. prob. & K \\ 
  \hline
$<$ 5 & 0.44 [0.43, 0.44] & 26885 &  &   0 & 0.47 [0.39, 0.55] &  55 & 0.95 [0.87, 1.03] &   3 \\ 
  5 - 9 & 0.47 [0.47, 0.48] & 23378 & 0.58 [0.37, 0.79] &  12 & 0.63 [0.57, 0.69] & 114 & 0.88 [0.80, 0.95] &  42 \\ 
  10 - 19 & 0.53 [0.53, 0.54] & 11683 & 0.71 [0.58, 0.84] &  33 & 0.82 [0.77, 0.87] & 159 & 0.87 [0.80, 0.95] &  42 \\ 
  20 - 29 & 0.59 [0.58, 0.60] & 2862 & 0.79 [0.69, 0.90] &  36 & 0.79 [0.71, 0.87] &  79 & 0.68 [0.51, 0.85] &  16 \\ 
  30 - 49 & 0.59 [0.57, 0.61] & 1720 & 0.77 [0.68, 0.87] &  58 & 0.83 [0.76, 0.91] &  74 & 0.84 [0.74, 0.93] &  38 \\ 
  50 - 99 & 0.76 [0.73, 0.78] & 713 & 0.76 [0.65, 0.87] &  52 & 0.83 [0.75, 0.91] &  68 & 0.86 [0.77, 0.96] &  31 \\ 
  100 - 299 & 0.51 [0.44, 0.58] & 142 & 0.81 [0.71, 0.90] &  59 & 0.91 [0.85, 0.98] &  51 & 0.80 [0.62, 0.97] &  17 \\ 
  $\geq$ 300 & 1.00 [1.00, 1.00] &   3 & 0.91 [0.84, 0.97] &  77 & 1.00 [1.00, 1.00] &   5 & 0.90 [0.71, 1.09] &  10 \\ 
  \hline
\end{tabular}
\end{table}

\begin{table}[H]
\caption{Median and interquartile range of standardized evidence inflation for for meta-analyses with different number of studies.}
\footnotesize	
\begin{tabular}{rrrrrrrrr}
  \hline
  & \multicolumn{2}{c}{Medicine} &  \multicolumn{2}{c}{Economics} &  \multicolumn{2}{c}{Psychology} & \multicolumn{2}{c}{Environmental} \\
  K & Evidence inflation / K & K & Evidence inflation / K & K & Evidence inflation / K & K \\ 
  \hline
$<$ 5 & 1.05 (1.00, 1.22) & 26885 &  &   0 & 1.11 (1.02, 1.42) &  55 & 1.69 (1.41, 1.91) &   3 \\ 
  5 - 9 & 1.05 (1.00, 1.22) & 23378 & 1.07 (1.03, 1.31) &  12 & 1.17 (1.04, 1.37) & 114 & 1.39 (1.24, 1.56) &  42 \\ 
  10 - 19 & 1.05 (1.01, 1.23) & 11683 & 1.28 (1.08, 1.51) &  33 & 1.36 (1.12, 1.56) & 159 & 1.29 (1.13, 1.55) &  42 \\ 
  20 - 29 & 1.06 (1.00, 1.27) & 2862 & 1.28 (1.10, 1.44) &  36 & 1.27 (1.05, 1.61) &  79 & 1.04 (1.01, 1.45) &  16 \\ 
  30 - 49 & 1.05 (1.00, 1.25) & 1720 & 1.19 (1.03, 1.35) &  58 & 1.41 (1.09, 1.75) &  74 & 1.21 (1.05, 1.68) &  38 \\ 
  50 - 99 & 1.14 (1.01, 1.40) & 713 & 1.23 (1.02, 1.37) &  52 & 1.31 (1.04, 1.69) &  68 & 1.16 (1.02, 1.24) &  31 \\ 
  100 - 299 & 1.01 (1.00, 1.20) & 142 & 1.15 (1.04, 1.32) &  59 & 1.47 (1.15, 1.80) &  51 & 1.12 (1.01, 1.23) &  17 \\ 
  $\geq$ 300 & 1.88 (1.64, 1.96) &   3 & 1.05 (1.02, 1.16) &  77 & 1.19 (1.18, 1.21) &   5 & 1.03 (1.02, 1.10) &  10 \\ 
   \hline
\end{tabular}
\end{table}

\begin{table}[H]
\caption{Mean and 95\% CI of standardized evidence inflation for for meta-analyses with different number of studies.}
\footnotesize	
\begin{tabular}{rrrrrrrrr}
  \hline
  & \multicolumn{2}{c}{Medicine} &  \multicolumn{2}{c}{Economics} &  \multicolumn{2}{c}{Psychology} & \multicolumn{2}{c}{Environmental} \\
  K & Evidence inflation / K & K & Evidence inflation / K & K & Evidence inflation / K & K \\ 
  \hline
$<$ 5 & 1.15 [1.15, 1.16] & 26885 &  &   0 & 1.20 [1.13, 1.27] &  55 & 1.63 [1.16, 2.29] &   3 \\ 
  5 - 9 & 1.15 [1.15, 1.15] & 23378 & 1.17 [1.05, 1.31] &  12 & 1.22 [1.18, 1.27] & 114 & 1.38 [1.31, 1.46] &  42 \\ 
  10 - 19 & 1.15 [1.15, 1.15] & 11683 & 1.29 [1.20, 1.38] &  33 & 1.34 [1.29, 1.38] & 159 & 1.32 [1.24, 1.41] &  42 \\ 
  20 - 29 & 1.16 [1.15, 1.17] & 2862 & 1.28 [1.21, 1.36] &  36 & 1.31 [1.25, 1.38] &  79 & 1.22 [1.08, 1.38] &  16 \\ 
  30 - 49 & 1.15 [1.14, 1.16] & 1720 & 1.23 [1.17, 1.30] &  58 & 1.41 [1.33, 1.50] &  74 & 1.33 [1.22, 1.45] &  38 \\ 
  50 - 99 & 1.21 [1.19, 1.23] & 713 & 1.24 [1.17, 1.31] &  52 & 1.36 [1.28, 1.45] &  68 & 1.18 [1.11, 1.25] &  31 \\ 
  100 - 299 & 1.13 [1.09, 1.16] & 142 & 1.18 [1.13, 1.23] &  59 & 1.42 [1.32, 1.52] &  51 & 1.13 [1.07, 1.19] &  17 \\ 
  $\geq$ 300 & 1.76 [1.43, 2.17] &   3 & 1.09 [1.07, 1.11] &  77 & 1.21 [1.14, 1.29] &   5 & 1.06 [1.02, 1.09] &  10 \\ 
   \hline
\end{tabular}
\end{table}

\begin{table}[H]
\caption{Median and interquartile range of bias of mean effect sizes for meta-analyses with different number of studies (in Cohen's $d$).}
\footnotesize	
\begin{tabular}{rrrrrrrrr}
  \hline
  & \multicolumn{2}{c}{Medicine} &  \multicolumn{2}{c}{Economics} &  \multicolumn{2}{c}{Psychology}  & \multicolumn{2}{c}{Environmental}\\
  K & Absolute bias & K & Absolute bias & K & Absolute bias & K \\  
  \hline
$<$ 5 & 0.10 (0.02, 0.22) & 26885 &  &   0 & 0.09 (0.03, 0.19) &  55 & 0.08 (0.07, 0.14) &   3 \\ 
  5 - 9 & 0.08 (0.00, 0.19) & 23378 & 0.07 (0.05, 0.32) &  12 & 0.07 (0.02, 0.11) & 114 & 0.15 (0.05, 0.27) &  42 \\ 
  10 - 19 & 0.07 (0.00, 0.17) & 11683 & 0.07 (0.03, 0.16) &  33 & 0.07 (0.03, 0.14) & 159 & 0.16 (0.06, 0.52) &  42 \\ 
  20 - 29 & 0.07 (0.00, 0.17) & 2862 & 0.09 (0.05, 0.18) &  36 & 0.03 (0.01, 0.13) &  79 & 0.22 (0.05, 0.54) &  16 \\ 
  30 - 49 & 0.08 (0.01, 0.18) & 1720 & 0.06 (0.02, 0.16) &  58 & 0.06 (0.03, 0.13) &  74 & 0.12 (0.04, 0.33) &  38 \\ 
  50 - 99 & 0.12 (0.03, 0.21) & 713 & 0.04 (0.01, 0.16) &  52 & 0.05 (0.01, 0.17) &  68 & 0.14 (0.07, 0.45) &  31 \\ 
  100 - 299 & 0.15 (0.03, 0.21) & 142 & 0.09 (0.01, 0.31) &  59 & 0.18 (0.06, 0.34) &  51 & 0.03 (-0.01, 0.16) &  17 \\ 
  $\geq$ 300 & 0.08 (0.06, 0.10) &   3 & 0.09 (0.03, 0.21) &  77 & 0.20 (0.04, 0.27) &   5 & 0.20 (0.04, 0.78) &  10 \\ 
  \hline
\end{tabular}
\end{table}

\begin{table}[H]
\caption{Mean and 95\% CI of bias of mean effect sizes for meta-analyses with different number of studies (in Cohen's $d$).}
\footnotesize	
\begin{tabular}{rrrrrrrrr}
  \hline
  & \multicolumn{2}{c}{Medicine} &  \multicolumn{2}{c}{Economics} &  \multicolumn{2}{c}{Psychology}  & \multicolumn{2}{c}{Environmental}\\
  K & Absolute bias & K & Absolute bias & K & Absolute bias & K \\  
  \hline
$<$ 5 & 0.14 [0.14, 0.14] & 26885 &  &   0 & 0.13 [0.09, 0.17] &  55 & 0.11 [0.02, 0.20] &   3 \\ 
  5 - 9 & 0.12 [0.11, 0.12] & 23378 & 0.17 [0.07, 0.28] &  12 & 0.09 [0.07, 0.12] & 114 & 0.19 [0.10, 0.28] &  42 \\ 
  10 - 19 & 0.11 [0.11, 0.11] & 11683 & 0.13 [0.07, 0.20] &  33 & 0.11 [0.09, 0.13] & 159 & 0.33 [0.21, 0.45] &  42 \\ 
  20 - 29 & 0.11 [0.10, 0.12] & 2862 & 0.15 [0.09, 0.20] &  36 & 0.09 [0.04, 0.14] &  79 & 0.34 [0.16, 0.52] &  16 \\ 
  30 - 49 & 0.13 [0.11, 0.14] & 1720 & 0.12 [0.08, 0.16] &  58 & 0.13 [0.08, 0.17] &  74 & 0.31 [0.17, 0.45] &  38 \\ 
  50 - 99 & 0.26 [0.22, 0.30] & 713 & 0.11 [0.07, 0.16] &  52 & 0.13 [0.08, 0.17] &  68 & 0.40 [0.16, 0.64] &  31 \\ 
  100 - 299 & 0.14 [0.10, 0.18] & 142 & 0.20 [0.12, 0.28] &  59 & 0.27 [0.16, 0.38] &  51 & 0.20 [-0.02, 0.43] &  17 \\ 
  $\geq$ 300 & 0.09 [0.04, 0.13] &   3 & 0.17 [0.12, 0.21] &  77 & 0.40 [-0.13, 0.92] &   5 & 0.95 [-0.26, 2.15] &  10 \\ 
  \hline
\end{tabular}
\end{table}

\begin{table}[H]
\caption{Median and interquartile range of overestimation factor of mean effect sizes for meta-analyses with different number of studies.}
\footnotesize	
\begin{tabular}{rrrrrrrrr}
  \hline
  & \multicolumn{2}{c}{Medicine} &  \multicolumn{2}{c}{Economics} &  \multicolumn{2}{c}{Psychology}  & \multicolumn{2}{c}{Environmental}\\
  K & Overestimation factor & K & Overestimation factor & K & Overestimation factor & K \\ 
  \hline
$<$ 5 & 1.66 [1.63, 1.70] & 26885 &  &   0 & 1.66 [0.99, 2.33] &  55 & 1.17 [0.98, 1.37] &   3 \\ 
  5 - 9 & 1.56 [1.53, 1.60] & 23378 & 1.65 [0.19, 3.11] &  12 & 1.37 [1.05, 1.69] & 114 & 1.31 [1.03, 1.60] &  42 \\ 
  10 - 19 & 1.56 [1.51, 1.61] & 11683 & 1.68 [0.79, 2.57] &  33 & 1.30 [1.09, 1.51] & 159 & 1.57 [1.12, 2.02] &  42 \\ 
  20 - 29 & 1.58 [1.48, 1.68] & 2862 & 1.86 [1.02, 2.70] &  36 & 1.28 [0.94, 1.61] &  79 & 2.02 [0.64, 3.41] &  16 \\ 
  30 - 49 & 1.75 [1.59, 1.92] & 1720 & 1.58 [0.95, 2.21] &  58 & 1.38 [0.99, 1.78] &  74 & 2.09 [1.01, 3.18] &  38 \\ 
  50 - 99 & 2.66 [2.00, 3.33] & 713 & 1.64 [0.95, 2.33] &  52 & 1.41 [0.96, 1.86] &  68 & 3.57 [-1.56, 8.71] &  31 \\ 
  100 - 299 & 2.61 [1.37, 3.85] & 142 & 3.45 [1.41, 5.50] &  59 & 1.58 [0.81, 2.35] &  51 & 1.41 [0.50, 2.32] &  17 \\ 
  $\geq$ 300 & 1.19 [0.43, 1.95] &   3 & 31.40 [-196.65, 259.45] &  77 & -5.69 [-54.48, 43.09] &   5 & -4.00 [-18.89, 10.90] &  10 \\  
  \hline
\end{tabular}
\end{table}

\begin{table}[H]
\caption{Mean and 95\% CI of overestimation factor of mean effect sizes for meta-analyses with different number of studies.}
\footnotesize	
\begin{tabular}{rrrrrrrrr}
  \hline
  & \multicolumn{2}{c}{Medicine} &  \multicolumn{2}{c}{Economics} &  \multicolumn{2}{c}{Psychology}  & \multicolumn{2}{c}{Environmental}\\
  K & Overestimation factor & K & Overestimation factor & K & Overestimation factor & K \\ 
  \hline
$<$ 5 & 1.32 (0.57, 2.06) & 26885 &  &   0 & 1.32 (0.97, 1.73) &  55 & 1.12 (1.10, 1.25) &   3 \\ 
  5 - 9 & 1.24 (0.56, 1.91) & 23378 & 1.31 (0.99, 1.76) &  12 & 1.20 (1.03, 1.55) & 114 & 1.25 (1.06, 1.40) &  42 \\ 
  10 - 19 & 1.23 (0.68, 1.89) & 11683 & 1.21 (0.88, 1.90) &  33 & 1.16 (1.03, 1.39) & 159 & 1.24 (1.07, 1.61) &  42 \\ 
  20 - 29 & 1.20 (0.68, 1.82) & 2862 & 1.54 (1.17, 2.37) &  36 & 1.10 (1.02, 1.41) &  79 & 1.24 (0.64, 1.53) &  16 \\ 
  30 - 49 & 1.25 (0.61, 1.99) & 1720 & 1.19 (1.07, 1.91) &  58 & 1.13 (1.05, 1.58) &  74 & 1.24 (1.05, 1.73) &  38 \\ 
  50 - 99 & 1.16 (0.57, 1.73) & 713 & 1.14 (0.99, 1.76) &  52 & 1.10 (1.00, 1.37) &  68 & 1.16 (0.44, 2.17) &  31 \\ 
  100 - 299 & 1.22 (-2.60, 4.19) & 142 & 1.09 (-0.41, 2.06) &  59 & 1.16 (0.90, 1.41) &  51 & 1.05 (0.95, 1.17) &  17 \\ 
  $\geq$ 300 & 1.20 (1.16, 1.24) &   3 & 1.10 (-1.17, 2.81) &  77 & 1.12 (1.09, 2.35) &   5 & 1.18 (-0.24, 2.02) &  10 \\ 
  \hline
\end{tabular}
\end{table}

\begin{table}[H]
\caption{Median and interquartile range of posterior probability of the presence of publication selection bias for meta-analyses with different number of studies.}
\footnotesize	
\begin{tabular}{rrrrrrrrr}
  \hline
  & \multicolumn{2}{c}{Medicine} &  \multicolumn{2}{c}{Economics} &  \multicolumn{2}{c}{Psychology}  & \multicolumn{2}{c}{Environmental}\\
  K & Post. prob. & K & Post. prob. & K & Post. prob. & K \\ 
  \hline
$<$ 5 & 0.49 (0.38, 0.59) & 26885 &  &   0 & 0.55 (0.48, 0.63) &  55 & 0.50 (0.46, 0.54) &   3 \\ 
  5 - 9 & 0.48 (0.37, 0.63) & 23378 & 0.55 (0.46, 0.74) &  12 & 0.57 (0.43, 0.67) & 114 & 0.58 (0.43, 0.67) &  42 \\ 
  10 - 19 & 0.50 (0.35, 0.71) & 11683 & 0.72 (0.47, 0.88) &  33 & 0.58 (0.40, 0.72) & 159 & 0.56 (0.45, 0.71) &  42 \\ 
  20 - 29 & 0.51 (0.32, 0.76) & 2862 & 0.85 (0.67, 0.96) &  36 & 0.49 (0.33, 0.75) &  79 & 0.63 (0.56, 0.77) &  16 \\ 
  30 - 49 & 0.52 (0.31, 0.86) & 1720 & 0.70 (0.45, 0.96) &  58 & 0.58 (0.42, 0.83) &  74 & 0.66 (0.37, 0.93) &  38 \\ 
  50 - 99 & 0.69 (0.40, 0.92) & 713 & 0.55 (0.24, 0.97) &  52 & 0.68 (0.30, 0.92) &  68 & 0.64 (0.26, 0.97) &  31 \\ 
  100 - 299 & 0.89 (0.33, 0.99) & 142 & 0.94 (0.28, 1.00) &  59 & 0.77 (0.30, 0.97) &  51 & 0.19 (0.14, 0.37) &  17 \\ 
  $\geq$ 300 & 0.73 (0.68, 0.83) &   3 & 0.99 (0.40, 1.00) &  77 & 0.39 (0.38, 0.69) &   5 & 0.95 (0.78, 1.00) &  10 \\ 
   \hline
\end{tabular}
\end{table}

\begin{table}[H]
\caption{Mean and 95\% CI of posterior probability of the presence of publication selection bias for meta-analyses with different number of studies.}
\footnotesize	
\begin{tabular}{rrrrrrrrr}
  \hline
  & \multicolumn{2}{c}{Medicine} &  \multicolumn{2}{c}{Economics} &  \multicolumn{2}{c}{Psychology}  & \multicolumn{2}{c}{Environmental}\\
  K & Post. prob. & K & Post. prob. & K & Post. prob. & K \\ 
  \hline
$<$ 5 & 0.50 [0.50, 0.51] & 26885 &  &   0 & 0.56 [0.52, 0.59] &  55 & 0.50 [0.40, 0.59] &   3 \\ 
  5 - 9 & 0.51 [0.51, 0.52] & 23378 & 0.61 [0.49, 0.73] &  12 & 0.57 [0.53, 0.60] & 114 & 0.56 [0.51, 0.61] &  42 \\ 
  10 - 19 & 0.54 [0.53, 0.54] & 11683 & 0.67 [0.58, 0.75] &  33 & 0.58 [0.54, 0.61] & 159 & 0.60 [0.54, 0.65] &  42 \\ 
  20 - 29 & 0.55 [0.54, 0.56] & 2862 & 0.79 [0.72, 0.86] &  36 & 0.54 [0.48, 0.59] &  79 & 0.65 [0.54, 0.76] &  16 \\ 
  30 - 49 & 0.57 [0.56, 0.58] & 1720 & 0.68 [0.60, 0.75] &  58 & 0.61 [0.55, 0.67] &  74 & 0.65 [0.55, 0.74] &  38 \\ 
  50 - 99 & 0.65 [0.63, 0.67] & 713 & 0.58 [0.49, 0.67] &  52 & 0.60 [0.53, 0.68] &  68 & 0.61 [0.48, 0.74] &  31 \\ 
  100 - 299 & 0.70 [0.64, 0.76] & 142 & 0.67 [0.57, 0.78] &  59 & 0.65 [0.55, 0.75] &  51 & 0.31 [0.17, 0.45] &  17 \\ 
  $\geq$ 300 & 0.77 [0.60, 0.94] &   3 & 0.73 [0.61, 0.85] &  77 & 0.58 [0.19, 0.98] &   5 & 0.75 [0.50, 1.00] &  10 \\ 
   \hline
\end{tabular}
\end{table}